\documentclass[sigconf]{acmart}
\usepackage[utf8]{inputenc}

\usepackage{graphicx}				
\usepackage{float}
\usepackage{url}


\copyrightyear{2022} 
\acmYear{2022} 
\setcopyright{rightsretained}\acmConference[MODELS '22 Companion]{ACM/IEEE 25th International Conference on Model Driven Engineering Languages and Systems}{October 23--28, 2022}{Montreal, QC, Canada}
\acmBooktitle{ACM/IEEE 25th International Conference on Model Driven Engineering Languages and Systems (MODELS '22 Companion), October 23--28, 2022, Montreal, QC, Canada}
\acmPrice{15.00}
\acmDOI{10.1145/3550356.3563131}
\acmISBN{978-1-4503-9467-3/22/10}

\begin{abstract}

    In this paper, 
    we propose a general architecture for designing language servers for hybrid modeling languages, that is, modeling languages that contain both textual and graphical representations.
    The architecture consists of a textual language server, a graphical language server, and a client that communicates with the two servers. The servers are implemented using the Language Server Protocol (LSP) and the Graphical Language Server Protocol (GLSP) and are based on a shared abstract syntax of the hybrid language.
    This means that only static resources need to be common between the graphical and textual language servers. The servers' separation allows each to be developed and maintained independently, while also enabling forward-compatibility with their respective dependencies.
    
    We describe a prototype implementation of our architecture in the form of a hybrid editor for the UML-RT language. 
    The evaluation of the architecture via this prototype gives us 
    useful insight into further generalization of the architecture and the way it is used.
    We then sketch a suitable extension of the architecture to enable support for multiple diagram types and, thus, multiple graphical views.

\end{abstract}

\begin{document}

\title{A General Architecture for Client-Agnostic Hybrid Model Editors as a Service}

\author{Liam Walsh}
\email{liam.walsh@queensu.ca}
\affiliation{
  \institution{Queen's University}
  \country{Canada}
}

\author{Juergen Dingel}
\email{dingel@queensu.ca}
\affiliation{
  \institution{Queen's University}
  \country{Canada}
}

\author{Karim Jahed}
\email{jahed@cs.queensu.ca}
\affiliation{
  \institution{Queen's University}
  \country{Canada}
}


    \maketitle

\section{Introduction}

With the increasing amount of complexity in software systems, it is important to be able to express them clearly and consistently. This is a must to ensure
effective communication among possibly diverse sets of stakeholders such as designers, developers, and users.
Abstracting away implementation-level details is one way to provide clarity to a user. Consistency can be achieved by a definition of rules and patterns. This type of approach to expressing systems allows them to be easily understood by way of familiarity with their domain, rather than programmatic knowledge.

The high-level motivation of our work is to enable better ways to express models of software as used in Model-Driven Development (MDD).
The ultimate goal is to facilitate the development of hybrid editors for modeling languages.

We propose a methodology of creating two separate services to equip Integrated Development Environments (IDEs) with the capabilities of hybrid editing.
This involves the implementation of two concrete syntaxes, 
one textual and the other graphical, each providing suitable representations (or views) of the same underlying model and its abstract syntax.
While this methodology seems straightforward conceptually, it is challenging to implement in a way that supports cohesion, maintainability and evolvability.

    \section{Background}

\subsection{Hybrid Modeling Languages}

Hybrid modeling languages are characterized by their use of more than one concrete syntax paradigm. In this paper, we will use the term to refer specifically to a modeling language which has both a textual concrete syntax, and a graphical concrete syntax.

\subsection{Language Servers} 

The classical method of implementing support for a language in an IDE usually results in high coupling to the IDE, even though most modern IDEs are functionally similar. Language servers take advantage of these similarities by treating IDEs as clients to be serviced and whose behaviour is dictated by the language's semantics. 

\subsubsection{Language Server Protocol}

The Language Server Protocol allows a client IDE to trade information and instructions with a language server. The language server can run as a background process and be queried by the IDE in order to give the user advanced editing features such as ``auto complete”, ``go to definition/declaration”, ``find all references”, among others \cite{langserver.org}. LSP allows for the client IDE to be completely language-agnostic. 

\subsubsection{Graphical Language Server Protocol (GLSP)}

The Graphical Language Server Protocol \cite{glsp-github, GLSPguide}, is an open source framework originally built to facilitate displaying SVG graphs and providing editing tools for graphical languages. This is all done in such a way that one implementation would be able to service multiple IDEs compliant with GLSP, hence the similarity in names between it and LSP.

GLSP functions by visiting a model, similar to how any other modeling tool would parse it, and systematically creating graph objects that reflect each different object in the original model specified by the developer. 
GLSP also allows the language developer to choose where to persist any elements of the concrete syntax that do not impact the execution semantics of the model (i.e. graph X/Y positioning, colors, shapes, etc.).

\subsection{UML for Real-Time}
The Unified Modeling Language for Real-Time (UML-RT) is a modeling language that is specified as a UML Profile. The language was created with the properties necessary to be able to describe Real-Time systems \cite{posse2016executable, BranSelic2002}. 
UML-RT's established graphical specifications made it a good candidate for our prototype hybrid editor. 

The two main components of a system's architecture are its \textit{structural} and \textit{behavioral} aspects. UML-RT is able to create graphical models of systems that are capable of specifying both these structural and behavioral aspects. Some tools for specifying UML-RT models, such as the open-source eclipse-based Papyrus-RT \cite{papyrusrtSpecs}, are able to automatically generate functional code from a given model. 

RTPoet is a toolset that we have created for the purposes of working with UML-RT models of different forms \cite{rtpoet}. The original use case was to bridge the gap between different UML-RT implementations (specifically Papyrus-RT \cite{papyrusrtSpecs} and RTIST/RSARTE \cite{rtist-hcl,rsarte-ibm}) by model transformation. We have already specified a textual implementation of UML-RT under the RTPoet umbrella. Once the UML-RT hybrid language server prototype is sufficiently mature, we intend to integrate it into the RTPoet suite.

    \section{Proposed Architecture}

In this section, we identify some key requirements that the architecture must satisfy. Then, the structure of the architecture is described together with sequence diagrams showing the interaction between components to realize aspects of these requirements.

The use of any language in an IDE must include the support of the standard CRUD (create, read, update, delete) operations. 
However, these are not a necessary point of focus for our proof-of-concept implementation, as these operations are mostly realized by the language definitions contained in the servers themselves.
We are more interested in the design of a high-level architecture.

Current best practices in GUI development include the use of the Model-View-Controller (MVC) design pattern. GLSP's design is informed by this pattern, as graphs naturally suggest the use of GUIs. We have chosen to model the textual language server in a similar way. While MVC is not normally applied to text editors, its properties allow us to maintain synchronization between views.
In the context of language servers, most of the responsibilities of the ``controller" and ``view" are carried out by the client, language-specific tasks are delegated to the server.

\subsection{Four Core Interface Requirements} 
\label{section:interfaces}

\begin{figure}[h]
    \centering
    \includegraphics[width=0.75\columnwidth]{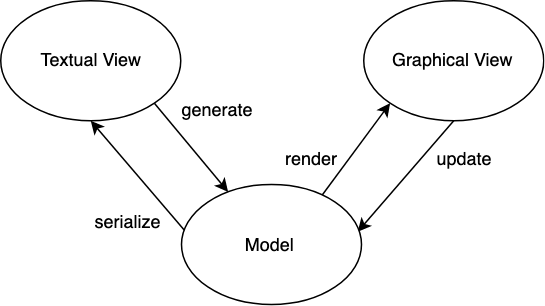}
    \caption{High-Level Relationship Between Graphical and Textual Views} 
    \label{fig:general_behaviour}
\end{figure}

Figure \ref{fig:general_behaviour} shows our proposed general approach to creating the architecture for hybrid language services. There are four main interfaces that must be implemented in order to fulfill the CRUD use cases in each of the textual and graphical views while keeping them synchronized. These four interfaces are denoted by the edges in the figure, and each represent a transformation to, or from, the views and model.
Essentially, the model is the abstract syntax, while the graphical and textual views are two different types of concrete syntaxes used to specify and represent it. The four different interfaces will be discussed in more detail below.

\subsubsection{Generation}

We will use ``generation" or ``regeneration" to refer to the transformation of the textual view into the model. This is the intended usage of Xtext \cite{EysholdtXtext2010}, and will not be any different than the first step of implementing a textual DSL (Domain-Specific Language) using the tool. The realization of this interface requires the specification of a grammar for the language. 
The textual language server uses the grammar to instruct the language client how to perform this transformation.

\subsubsection{Serialization}

The model that the textual view generates retains enough information about the original specification that it is still possible to perform a transformation from the model itself back to the textual view. This process is the inverse to the previously explained one, and the client is advised on how to do so by the textual language server.

\subsubsection{Rendering}

The ``rendering" of the graphical view should be realized by specifying a mapping between model objects and graph element objects (predominantly different types of nodes and edges). More granular parts of the language may be left out of this mapping, if they can not be appropriately graphed and can be more effectively handled by the textual view (e.g., action code in state machines).

\subsubsection{Updating}

Graphical views typically only display a subset of the elements of the model they represent.
Lower-level properties may be simplified or excluded altogether, meaning we cannot reliably retrieve these by transforming a graphical view into a model. 
We can, however, manipulate the model directly by keeping track of the graph elements' source mappings. 
GLSP allows for this to be done natively.
We will refer to this as ``updating" the model, as we are applying an update directly to the model objects by proxy of the graphical elements.
Models can be kept complete and correct
by restricting the operations available in the graphical view (often by using ``palettes of operations").
Each time an update is applied, the graphical view must then be rendered anew using the updated model.

\subsection{Architecture Diagram}

\begin{figure}[H]
    \centering
    \includegraphics[width=\columnwidth]{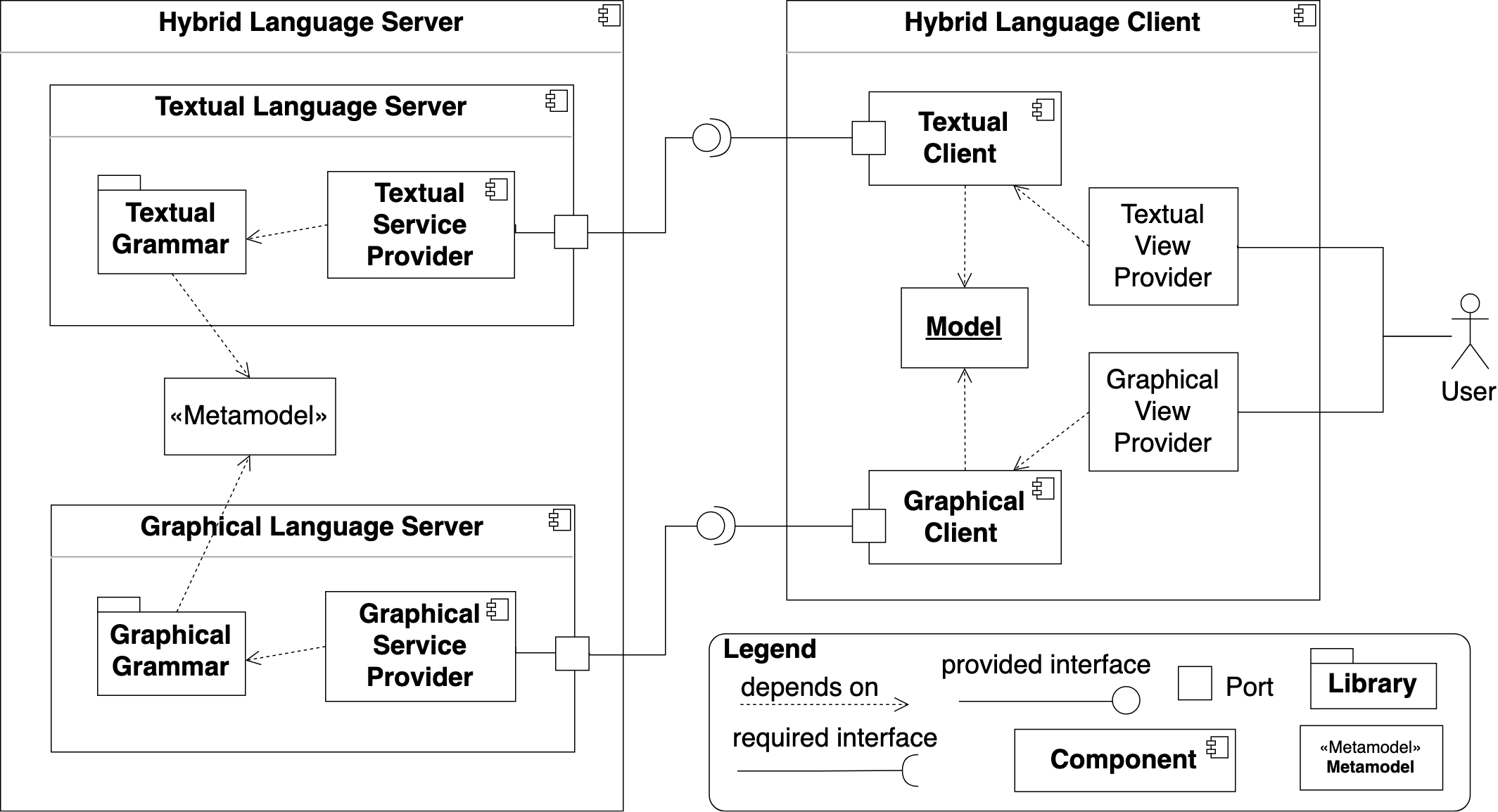}
    \caption{Architecture of Hybrid Language Server} 
    \label{fig:overall_architecture_original}
\end{figure}

Figure \ref{fig:overall_architecture_original} depicts our proposed general architecture for hybrid language servers. The diagram shows the communication between two language clients, encapsulated together in the ``hybrid language client", and two language servers, together in the ``hybrid language server". The two pairs of clients and servers communicate on different ports as, despite their similarities, they make use of different protocols.

The previously described ``generation" and ``serialization" occur between 
the ``textual client" and the ``model" shown in figure \ref{fig:overall_architecture_original}. 
Though, the methods that the client uses to do so are given to it by the textual language server. 
The same is true with respect to the graphical side: ``rendering" and ``updating" occur between the ``graphical client" and ``model", 
and the information on how to carry out these steps is held by the graphical language server.

\subsection{Sequence Diagrams}

In order to better illustrate the interaction between the modules in the proposed general architecture, some sequence diagrams depicting different behaviours have been included below.

\begin{figure}[H]
    \centering
    \includegraphics[width=\columnwidth]{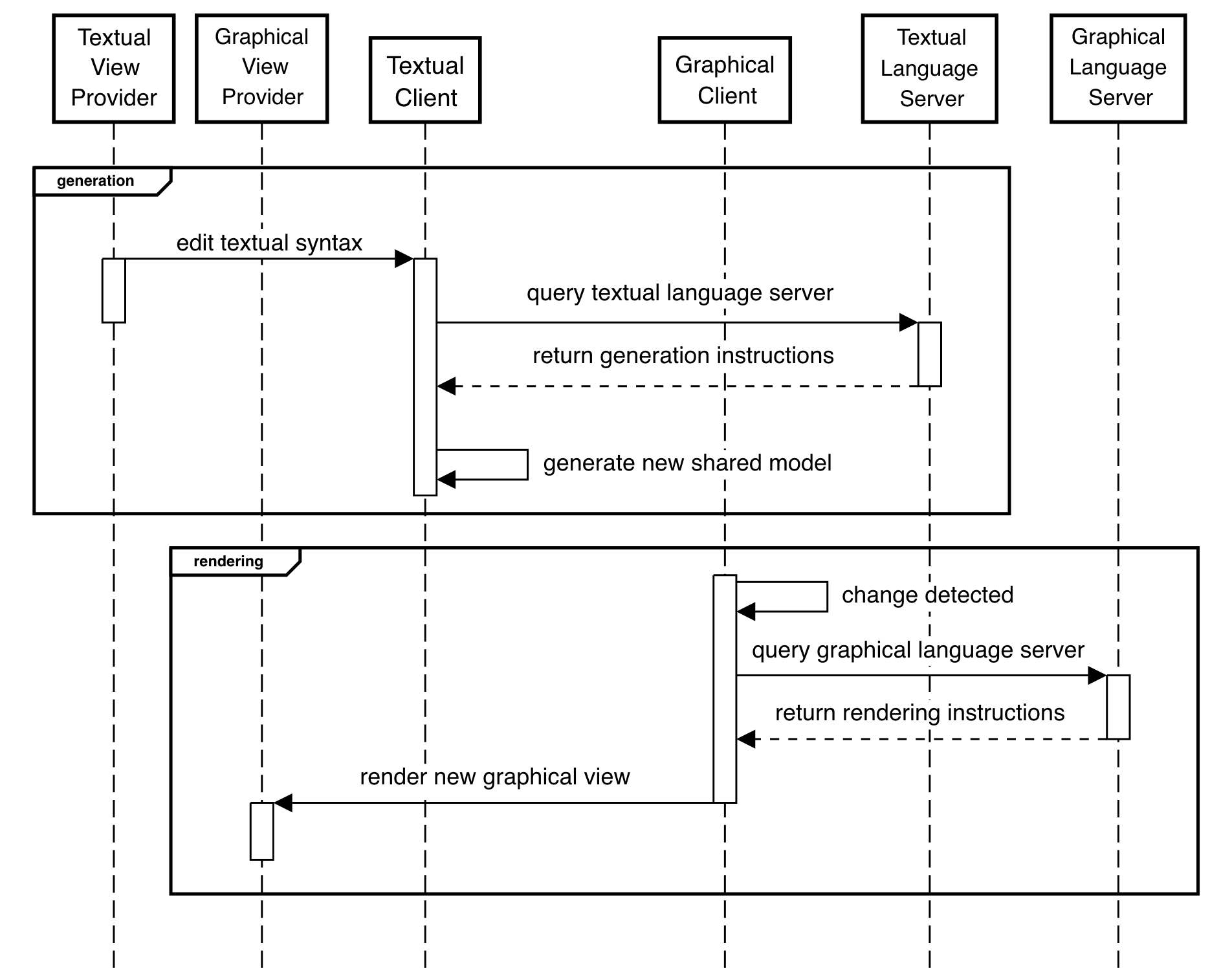}
    \caption{Sequence Diagram: Textual Edit Action}
    \label{fig:sequence_text_edit}
\end{figure}

The sequence diagram in figure \ref{fig:sequence_text_edit} shows how a change to the textual view would be propagated to the graphical view. 

\begin{figure}[H]
    \centering
    \includegraphics[width=\columnwidth]{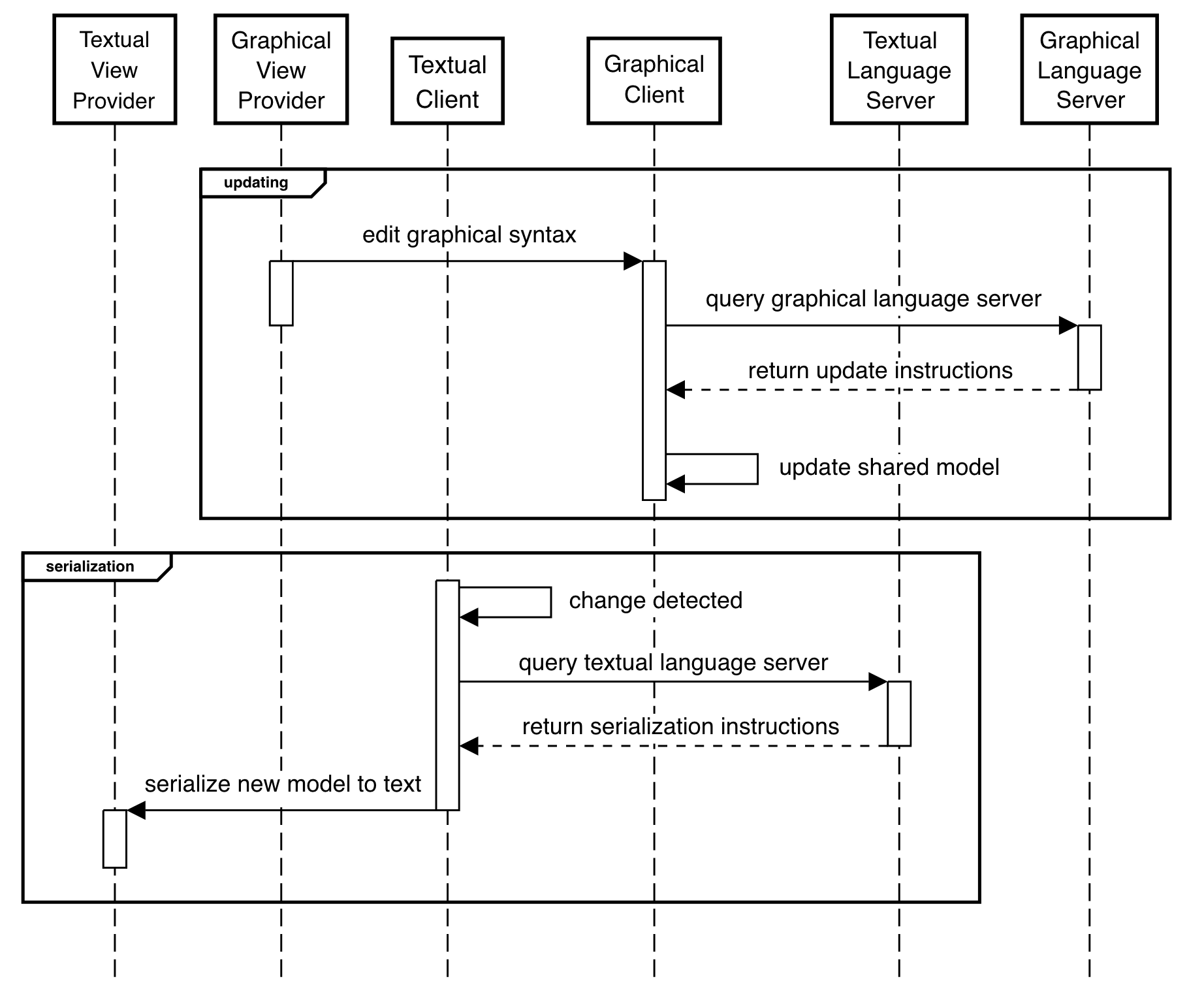}
    \caption{Sequence Diagram: Graphical Edit Action}
    \label{fig:sequence_graph_edit}
\end{figure}

The sequence diagram in figure \ref{fig:sequence_graph_edit} shows how a change to the graphical view would affect the textual view. By simply observing the shape of this sequence compared to the previous one, it is clear that they are very similar.

    \section{Prototype Implementation}

In order to evaluate our proposed architecture, we have created a prototype hybrid language server, which we will refer to as the ``RTPoet language", based on UML-RT. Below is a description of our implementation of the four core interfaces described in section \ref{section:interfaces}. This is followed by a discussion of our findings.

\subsection{Generation}

To implement the generation interface, a textual grammar for the RTPoet language has been specified using Xtext.

While most Xtext-based languages often define code generation instructions in addition to a grammar, the code stubs included when creating a new Xtext project are sufficient to allow model files to be generated from the textual view. This is possible because the model file is purely structural. 

We have made our Xtext implementation of UML-RT (i.e., the RTPoet language's textual syntax) available as part of the RTPoet tool suite, available on GitHub \cite{rtpoet-dsl}. 

\subsection{Serialization}

In order to be able to serialize a model back into the textual view, specific mappings to code templates must be specified for each of the model objects. Whereas the aforementioned generation process would typically yield ``markup language" string representations of objects as their target (such as EMF/XML models), this type of transformation would need to yield complete and correct ``code". The templates can consist of predefined code blocks with different types of values and identifiers substituted into the appropriate places. A traversal of the model's abstract syntax tree allows the string templates to be correctly ordered. This is similar to how an Xtext language might normally be equipped to generate code, but it easier to implement considering we already have access to the target language's grammar and metamodel.

\subsection{Rendering}
The metamodel resulting from the specification of the Xtext grammar was used to automatically create a full Java class model for the RTPoet language. The structure of these classes exactly reflects the scope of models specifiable in the textual syntax. This collection of Java classes was used as a basis for the implementation of the graphical language server. The collection's format is identical to the metamodel.
The next step was to map each metamodel type to an appropriate GLSP graphical type.
GLSP contains its own classes for graph elements, and assigns them SVG properties at runtime.
 
Next, a model visitor was created, based on one of GLSP's extension points, that converts any given model into a collection of graphical objects based on the mappings.
The graph resulting from our rendering operation is held directly in memory. The source mappings are kept track of by GLSP.
Since we have opted to not store any additional graphical properties, we chose to implement automatic arrangement of the elements. 

\subsection{Updating}

Some basic update instructions were added to the graphical language server, such as the ability to add a new simple state to a state machine. 
However, the implementation of the prototype was halted here, as there were architectural incompatibilities that needed to be addressed.
This will be discussed in detail below.

    \section{New Proposed Architecture}

We found that, given our current approach, the functionality of our implementation of UML-RT was not up to par with existing editors. 
In order to implement hybrid languages in a more convincing, user-friendly way, single views are insufficient. 
In particular, single views offer poor support for the following modeling concepts and activities.

\subsection{Containment}

A concept that is present in many modeling languages is that of object containment. If an object owns a collection of other objects, it may make sense to render the collection of objects within the boundaries of their owner. However, as object ownership depth increases, this becomes harder to manage.
A better solution is to render only objects with a direct containment relation. This would mean that in order to interact with elements of greater depth, we would need the ability to drill down into elements using new views.

\subsection{Projection}

Models often contain many elements, complicating their user-friendly graphical display. Projection, that is, the selective display of only specific subsets of model elements, can mitigate this problem.

\begin{figure}[H]
    \centering
    \includegraphics[width=\columnwidth]{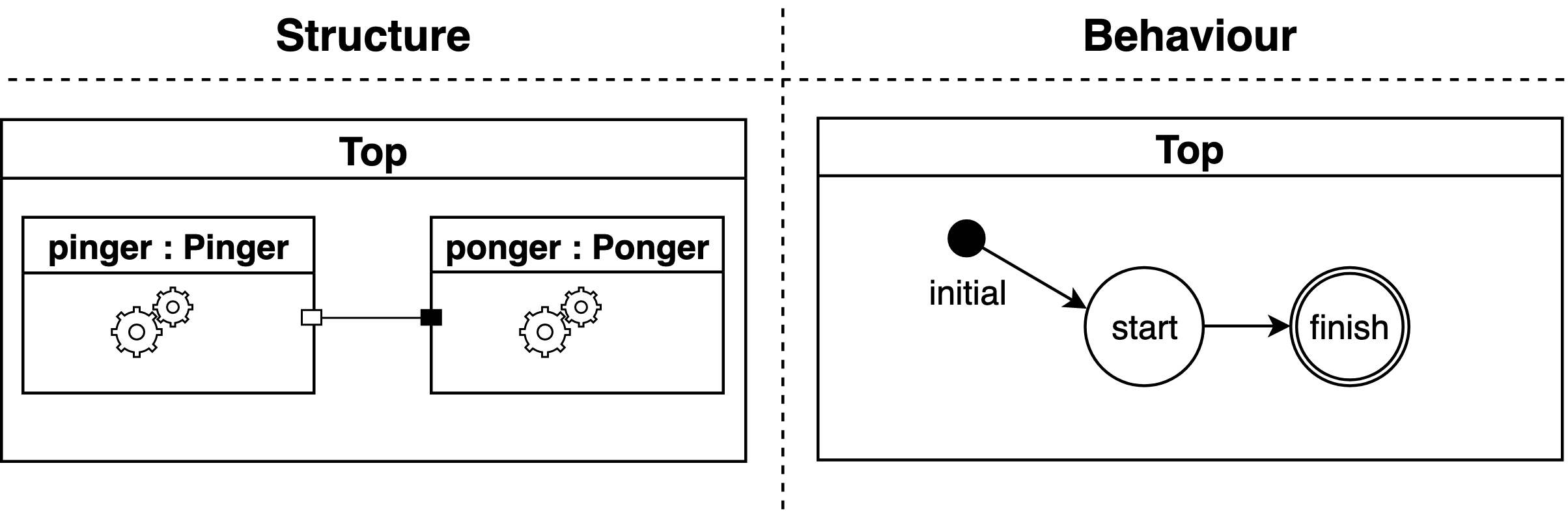}
    \caption{Structure vs Behaviour in UML-RT Model}
    \label{fig:structure_vs_behaviour}
\end{figure}

These projections can be based on language concepts, such as types or other fixed properties, and they can be further influenced by user input. For instance, some elements in a UML-RT model represent structure, while some other represent behaviour (as seen in Figure \ref{fig:structure_vs_behaviour}). While displaying different projections simultaneously in a single view can be useful, it can also cause confusion, and support for different views seems preferable in general.

\subsection{Static Analysis}

A static analysis often involves using the model as an input and, after some additional processing, returning an output whose syntax and semantics may well be outside the scope of the modeling language itself. For example, we might want to see a full traversal of a state machine. This would include duplicate states, something not possible using only the vanilla model view.
Therefore, rendering graphs for these types of operations is intrinsically incompatible with a single model visualization.

In summary, an architecture for hybrid language servers should enable support for different views. Only then can different parts of the model, as well as results of different projections and analyses, be displayed most appropriately.

By using multiple different graphical views, we may represent languages' structure more accurately while also providing an extension point for supplemental graph functionality. 

\subsection{Modified Architecture}

\begin{figure}[H]
    \centering
    \includegraphics[width=\columnwidth]{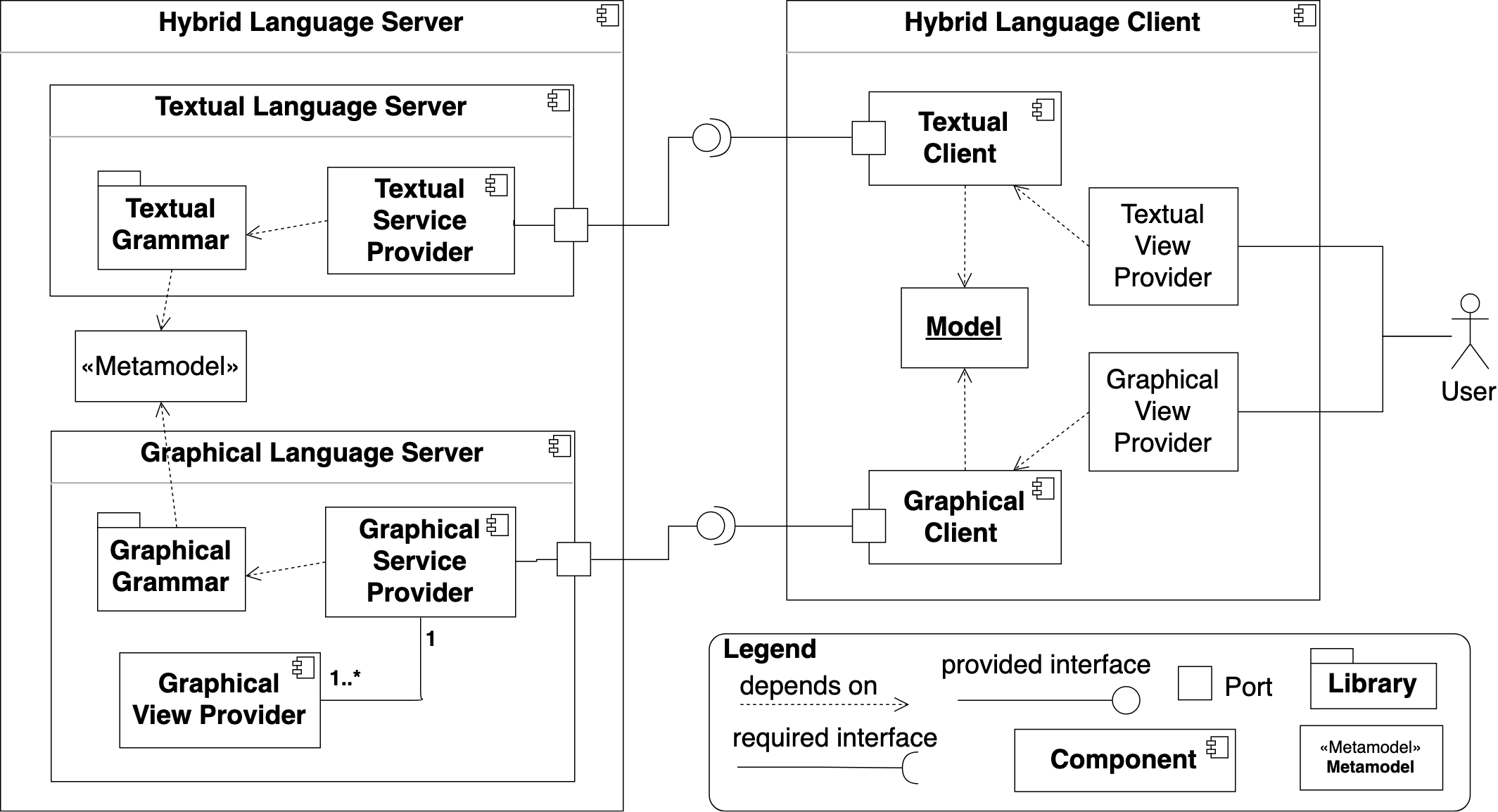}
    \caption{New Architecture of Hybrid Language Server}
    \label{fig:architecture}
\end{figure}

Our architecture can be modified to support multiple views. This modification is minimally invasive and, from the point of view of the client, it is the same as before. The new proposed architecture can be seen in figure \ref{fig:architecture}.

\subsubsection{Switching Views}

\begin{figure}[H]
    \centering
    \includegraphics[width=\columnwidth]{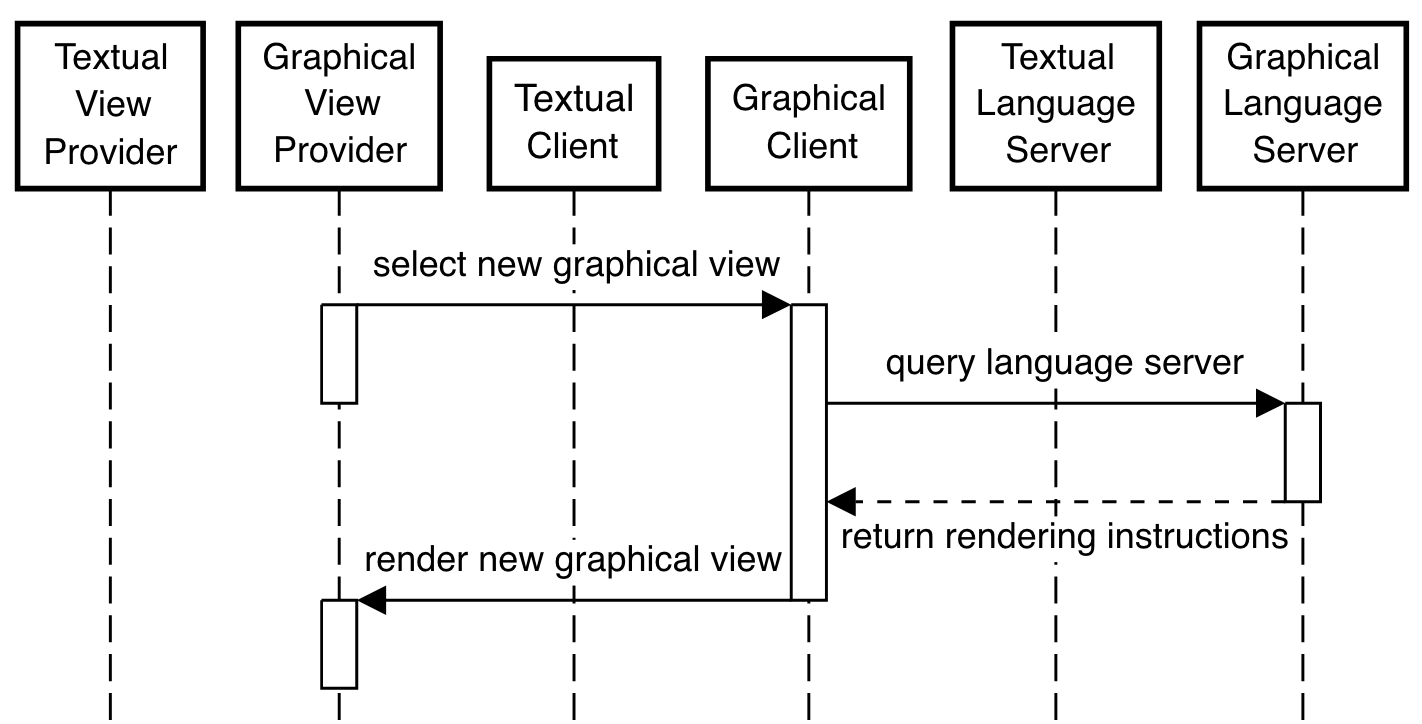}
    \caption{Sequence: Graphical View Switching}
    \label{fig:sequence_view_switch}
\end{figure}

A short sequence diagram depicting the switching of graphical views can be seen in figure \ref{fig:sequence_view_switch}. The initial action could come from simply clicking on an element. In this case, the client would send the click with a reference to the clicked element along with necessary metadata. We can also see that the latter half of the sequence is the same as the previous graph rendering sequence in figure \ref{fig:sequence_text_edit}. Unless we explore the local scope of the Graphical Language Server, the process will not appear much different.

    \section{Related Work}

\paragraph{Passing Graphical Information Using LSP}

Papers \cite{RodriguezMid2018} and \cite{RodriguezLate2018} discuss the proposal and implementation of a custom graphical editing solution for EMF languages, using LSP without any extensions. Their solution involves converting the EMF model into a JSON-based format in order to exchange it with the server across LSP. The JSON-based equivalent of the whole EMF model is what is transmitted, including pure graph properties to render the graph as an SVG (e.g. \texttt{width:100, height:50, shape:square}). The usage of JSON-based messages makes this effectively an implementation of a textual language server. They also define actions that the front-end client is able to perform and manually map them to specific LSP messages, such as creating or removing edges or nodes. 
In the earlier of the their two papers, they propose the solution as completely compliant with LSP as a prototype. In the later paper, they discuss the possibility of extend LSP or defining a brand new protocol, but ultimately decide to remain compliant with LSP. 
Since our work separates the two editing mediums' implementations, we will not need to evaluate problems such as protocol extensions.

\paragraph{UML Profile Hybrid Editors}
Addazi et al. \cite{ernestoProfileHybrid} propose a methodology of implementing hybrid languages based on UML by extending it (i.e. creating UML profiles). In concept, their approach is similar to ours, as it makes use of a shared abstract syntax between the two different views of a given language. Similar to our proposed architecture, the source of their shared metamodel is based on Xtext's grammar specifications. The authors use UML as a basis for their specification of Xtext grammars, allowing them to leverage its existing graphing libraries. Our approach does not rely on UML, but it could be extended in the same way (i.e. used as a basis for typing in the Xtext grammar specification). However, reusing any of the graphical editor implementation designs in this work is not possible in the context of language servers. The UML graphing libraries' dependence on IDE-specific features render them incompatible with our proposed architecture both in concept and in practice. That being said, the stability of UML could motivate the creation of similar UML-based graphing libraries that operate within the boundaries of graphical language servers.

\paragraph{bigER Modeling Language}
Glaser et al. have created the ``bigER" tool \cite{bigER} which, as of yet, seems to be the most closely related to our own work. The bigER tool is an example of a functional hybrid language server that also makes use of Xtext. The approach to the design of the bigER tool mirrors one of our prior proposed architectures for hybrid language servers \cite{walsh2020toward, walsh2020slides}. At a high level: graphical operations are mapped to code snippets which are injected directly into to the textual model. This is similar to the code templating required for defining the ``serialization" interface, but controlled by a palette of operations similar to that required by the ``updating" interface (as seen in figure \ref{fig:general_behaviour}). The textual model can then regenerate the graphical model, displaying the changes graphically. We did not pursue this style of architecture for several reasons. Most notably, this solution to synchronization of the two editors is more difficult to specify and maintain as the size of the language increases. For more complex languages, static analysis might need to be done on the code in order properly contextualize the code snippets. Our current proposed architecture uses the language's structure to guarantee completeness and correctness of graphical edits. Language specification errors in our implementation can more easily be detected at compile time, whereas errors using this style of implementation may only be clear at runtime.

\paragraph{Graphical Viewspoints}
In \cite{2007GraphViews}, domain-specific modeling languages with multiple ``viewpoints" are discussed. The work describes the nuances of the relationship between model and graphs. The authors present a methodology for designing multi-view DSVLs (Domain-Specific Visual Languages) in which they note that models can be projected into submodels in different ``viewpoints" (which are analogous to the ``views" described in this work). The concept of viewpoints closely mirrors our proposed implementation of views. While our work was not inspired by this paper, it corroborates our findings well. The authors' work predates LSP, meaning that it evidently does not make use of it. This is a key differentiating factor relative to our work.

    \section{Conclusion and Future Work}

We have proposed and evaluated a methodology for designing and implementing hybrid language servers, in the form of a general architecture, that offers a cohesive approach, simplifies maintainability, and facilitates language evolution. 

We anticipate that there is much work to be done in the study of hybrid language design and usability. This may be inspired by existing research on best practices for textual and graphical mediums alike \cite{physicsOfNotation, tufte1998visual}. 
We hope that our findings may be foundational to the accessibility of hybrid modeling languages, such that we may facilitate future work in studying them as a whole, and not only relative to language servers.

    \bibliographystyle{ACM-Reference-Format}
    \bibliography{main}

\end{document}